\documentclass{aa}
\usepackage{psfig}
\usepackage{epsf}
\usepackage{graphicx}

\newcommand{\Msun}{\mbox{$M_{\odot}\;$}}
\newcommand{\Lsun}{\mbox{$L_{\odot}\;$}}
\def\lsim{\;\raise0.3ex\hbox{$<$\kern-0.75em\raise-1.1ex\hbox{$\sim$}}\;}
\def\gsim{\;\raise0.3ex\hbox{$>$\kern-0.75em\raise-1.1ex\hbox{$\sim$}}\;}

\def\mathnew{\mathsurround=0pt}
\def\simov#1#2{\lower .5pt\vbox{\baselineskip0pt \lineskip-.5pt
        \ialign{$\mathnew#1\hfil##\hfil$\crcr#2\crcr\sim\crcr}}}

\def\cmc{\rm ~cm^{-3}}

\def \kms {\rm ~km~s^{-1}}

\def\ergs{\rm ~erg~s^{-1}}

\def\etal{{ et al. }}

\def \chan {{\it Chandra}}
\def \xmm {{\it XMM-Newton}}

\def \hcm {\hbox {\ifmmode $ atom cm$^{-2}\else atom cm$^{-2}$\fi}}
\def \arcmin {\hbox{$^\prime$}}

\def\approxgt{\mathrel{\hbox{\rlap{\lower.55ex \hbox {$\sim$}}
        \kern-.3em \raise.4ex \hbox{$>$}}}}
\def\approxlt{\mathrel{\hbox{\rlap{\lower.55ex \hbox {$\sim$}}
        \kern-.3em \raise.4ex \hbox{$<$}}}}
\begin{document}

%\thesaurus{(02.01.1; 02.18.5; 09.03.1; 09.03.2; 09.09.1: \src; 09.19.2)}

%\title{Numerous Faint Hard X-ray Sources in the Galactic Center:Supernova Ejecta Fragments Contribution}

\title{Faint Hard X-ray Sources in the Galactic Center Region: \\ Supernova
Ejecta Fragments Population}

\titlerunning{Hard X-ray Sources in the Galactic Center}

%\end{center}

\author{A. M. Bykov
}

\institute{A.F. Ioffe Institute for Physics and Technology,
          St. Petersburg, 194021, Russia}

\offprints{e-mail: byk$@$astro.ioffe.ru}

\date{Received  }

\abstract{Long \chan\ observations of $17 \arcmin \times 17
\arcmin$ field in the Galactic Center (GC) region reported by Muno
\etal. (2003) have revealed a population of about 2,000 hard faint
X-ray sources limited by luminosities $L_{\rm x} \geq$ 10$^{31}~
\ergs$. We show that fast moving knots (FMKs) of supernovae ejecta
could comprise a sizeable fraction of the sources. Each supernova
event can produce hundreds of FMKs. The presence of $\sim 3$
supernova remnants of age $\sim 10^3$ ${\rm years}$ in the GC
region could provide the required number of the FMKs. Simulated
X-ray spectra of the FMKs contain both thermal and nonthermal
components. The nonthermal spectra are power-laws with hard photon
indexes $0 \leq \Gamma \leq 1.5$ and prominent lines of some
metals (e.g. Fe).
%providing hardness ratios close to that observed in the \chan\ sample.
The ${\rm log}$N--${\rm log}$S distribution and the hardness
ratios of the FMKs in the GC region  are consistent with the
\chan\ findings. Spatial distribution of the
%point-like bright
FMKs in the GC region should appear as an
ensemble of coherent shell and jet-like structures of a few
$arcmin$-scale.
 }

\maketitle

\section{Introduction}
The unique activity of the GC appears in a number of spectacular
phenomena observed in the multi-wavelength studies of the region.
Apart from the enigmatic central compact source Sgr A$^{\ast}$,
there are early-type stars with more than two dozens of blue
supergiants, providing the luminosity of the central parsec of
$\sim$ 10$^8\, \Lsun$, supernova remnants (SNRs), dense molecular
clouds, extended magnetic structures, evidences for high
star-forming activity in the past (a few 10$^7$ ${\rm years}$ ago)
(e.g. Blitz et al. 1993; Mezger, Duschl \& Zylka 1996).

X-ray observations of the GC region with sensitive high-resolution
 telescopes \chan\ and $\xmm\ \,$ are a powerful tool to
study the rich but highly obscured GC region. \chan\ observations
were performed for %the Sgr B2 (Murakami \etal 2001),
the Sgr A East (Maeda \etal 2002) and the GC regions ( Wang,
Gotthelf \& Lang 2002; Baganoff et al 2003;  Muno et al. 2003).

With a deep 590 ks \chan\ exposure of a $17 \arcmin \times 17
\arcmin$ field in the GC region Muno \etal (2003) (M03 hereafter)
catalogued 2357 X-ray sources limited by the luminosity $L_{\rm x}
\geq$ 10$^{31}~ \ergs$ (2.0--8.0 ${\rm keV}$).
%A substantial part of the \chan\ sample sources
%positions are projected into the Circum-Nuclear Disc (Mezger \etal
%1996) between 1.7 and 7 ${\rm pc}$ around the GC.
% and containing $\sim$ 10$^4 \Msun$ of highly clumped matter.
%This imply the average number density in the CND ~ 270 \cmc.
The ${\rm log}$N--${\rm log}$S distribution of the GC sources is
steep around the sample limiting luminosity, indicating that the
weak point sources could contribute substantially to the observed
X-ray diffuse emission. The problem of the origin of the observed
large scale X-ray emission from the Galactic ridge requires a
careful study of possible classes of abundant hard X-ray sources
with $L_{\rm x} \gsim 10^{29} \ergs$ (e.g. Tanaka, Miyaji \&
Hasinger 1999). Because of the hard spectra of the detected
sources M03 suggested that accreting magnetic white dwarfs (or
magnetic cataclysmic variables) and wind accreting neutron stars
could be possible candidates for the sample. Moreover, Muno \etal
(2003a) found periodic variability for eight sources in a sample
of 285 sources and suggested that more sources in the sample could
be magnetic white dwarfs or accreting pulsars. It is not yet
clear, however, how many sources of the both classes are present
in the GC region (M03). We shall consider here another potentially
abundant class of hard X-ray sources related to supernova activity
in the dense GC medium.

Multiwavelength studies of SNRs have revealed a complex structure
of metal ejecta with the presence of  fast moving isolated
fragments of SN ejecta, interacting with the surrounding media. In
the optical the multiple fast moving knots (FMKs) were observed
outside the main shell of Cas A (e.g. Chevalier \& Kirshner 1979;
Fesen \etal 2002) and in some other SNRs. Optical FMKs in Cas A
are very abundant in O-burning and Si-group elements. They have a
broad velocity distribution around 6,000 $\kms$ and apparent sizes
below 0.01 ${\rm pc}$. The optical knots in Cas A should have
$L_{\rm x} \leq 10^{29} \ergs$ because of low ambient density. The
similar FMKs in the dense environment of the GC region would have
$L_{\rm x} \geq 10^{30} \ergs$ with bright IR counterparts (Bykov
2002). In a low density medium only relatively big FMKs can be
observed in X-rays, like the Vela shrapnel A. \chan\ and \xmm\
observations revealed there a head-tail structure with a prominent
Si line, indicating that the object is a fast ejecta fragment of
the scale $\sim$ 0.3 ${\rm pc}$ (e.g. Aschenbach (2002)). Some of
the hard X-ray sources detected with \xmm\ in IC~443 are likely to
be fragments of the SN ejecta of size below 0.1 ${\rm pc}$
interacting with a molecular cloud border (Bocchino \& Bykov
2003).

In the present paper we show that an ensemble of hard X-ray
sources associated with fast moving supernova ejecta fragments
could be abundant and can account for some of the observed
properties of the detected GC sources such as the hardness ratios
and the ${\rm log}$N--${\rm log}$S distribution. We simulate X-ray
spectra of the FMKs of different velocities in the GC environment.
In our model the hard X-ray emission of FMKs is due to both hot
thermal postshock plasma and nonthermal particles accelerated at
the bow shock.

\section{X-ray emission of FMKs in the Galactic Center region}
Consider an FMK propagating through the intercloud gas which
pervades the central 50 ${\rm pc}$. Mezger \etal (1996) estimated
the gas average density of some 10$^2 \cmc$ and the mass of atomic
hydrogen $\sim$ 10$^6 \Msun$, i.e. about one half of the total
mass in this region. The matter is highly clumped. Magnetic field
in the GC region is of mG strength, and provides a substantial
contribution to the total pressure.

The lifetime of an FMK in a dense media is an important factor for
our model and should be considered first. A fast moving knot is
decelerating due to the interaction with the ambient gas. The drag
deceleration time of a knot of velocity $v$, mass ${\cal M}$ and
radius ${\cal R}$ can be estimated as $\tau_{\rm d} \sim {\cal M}/
(\rho_{\rm a} v \pi {\cal R}^2) \approx 10^3 \cdot {\cal M}_{\rm
-3}/(n_{\rm a2}\, v_{\rm 8}\,{\cal R}_{\rm -2}^{2}\,)$ ${\rm
years}$. Here ${\cal M}_{\rm -3} = {\cal M}/10^{-3}\,\Msun$ and
${\cal R}_{\rm -2} = {\cal R}/(0.01 ${\rm pc}$)$. The number
density $n_{\rm a2}$ of the ambient matter is measured in 100
$\cmc$ and the FMK velocity $v_{\rm 8}$ is measured in 1,000
$\kms$. In the inner 20$\arcmin$ of the GC region the average
number density $n_{\rm a} \gsim 100 \cmc$ (e.g. Mezger \etal 1996)
and the fragment deceleration time $\tau_{\rm d} \lsim 10^3 {\rm
years}$.

The high pressure gas in the head of the fast moving ejecta
fragment could  drive an internal shock resulting in the knot
crush and fragmentation. The internal shock velocity $v_{\rm is}
\approx v/\chi^{1/2}$, where the density contrast $\chi =
\rho_{\rm k}/\rho_{\rm a}$, and $\rho_{\rm a}$ and $\rho_{\rm k}$
are the ambient gas and the dense fragment densities,
respectively. The knot crushing time scale $\tau_{\rm c} = {\cal
R}/v_{\rm is}$. There are 2D hydrodynamical simulations showing
that the fast knot fragmentation occurs on the timescale of $\sim
(3-4) \tau_{\rm c}$ (e.g. Klein \etal 1994; Wang \& Chevalier
2002). Wang \& Chevalier (2002) studied the effect of density profile for
core-collapsed SNe  to produce the protrusions in the
outer shock  as observed in the Vela SNR.
 For the FMKs of mass  ${\cal M}_{\rm -3} \sim 1$ and
${\cal R}_{\rm -2} \sim 1$  the fragmentation time scale is a
factor of $\sim$ 2.5 less than the drag deceleration time.
However, the FMK will be a source of hard X-ray
emission even at the fragmented stage if the strong forward shock
is still driven by the FMK.

The hydrodynamical estimation of the inner shock velocity
$v_{\rm is}$ given above assumes an efficient conversion of the
bow shock ram pressure to the knot internal shock. The effects of
 nonthermal particle acceleration reduce the postshock gas pressure
and the internal shock velocity thus increasing $\tau_{\rm c}$
(Bykov 2002). In that case the life time would be close to
$\tau_{\rm d}$, unless the ablation processes. The effect of magnetic fields
and nonthermal particles on the knot ablation is not yet studied.

Massive star winds would change the circumstellar environment,
creating caverns of low density matter. The lifetime of fast
fragments of a SN in the low density cavern should be longer than
$R_{\rm b}/v_{\rm k}$, where $R_{\rm b}$ is a wind bubble size.
Following the approach by Chevalier (1999) and accounting for the
high pressure of the GC ambient matter one may estimate the
typical scale $R_{\rm b} \sim$ 3 ${\rm pc}$, for WR, while $R_{\rm
b}$ is $\sim$ 1 ${\rm pc}$ for O9-B0 progenitor stars. The
lifetimes of FMKs are $\sim 10^3$ ${\rm years}$. The collective
effect of powerful stellar winds in compact associations could
blow out more extended caverns, providing somewhat longer
lifetimes for the fragments of the SNe exploded inside the cavern.

Supersonic motion of the FMKs in the ambient medium result in a
bow-shock/knot-shock structure heating postshock plasma and
accelerating fast particles. Both thermal and nonthermal emission
of postshock plasma can produce X-ray photons.

\subsection{The thermal emission of FMKs}

Thermal emission of hot postshock plasma could be observed from
some FMKs even from the highly obscured GC region. The postshock
ion temperature (measured in 10$^7$ K) can be estimated as $T_{\rm
i7} \approx 1.4~v_{\rm 8}^2$ for the simplest single-fluid case of
a plasma of the solar abundance. The electron temperature just
behind the strong shock is typically much lower. In the postshock
layer the electron and ion temperatures equilibrate due to Coulomb
collisions. The complete e-i Coulomb equilibration requires the
time $t_{\rm ei} \gsim 3.2 \cdot 10^{11}~T_{\rm i7}^{3/2}/n~(s)$,
where the postshock density $n$ is in $\cmc$ (e.g. Mewe 1990). The
maximal electron temperature in the postshock region can be
estimated from the relation $\tau_{\rm d} \geq t_{\rm ei}$
providing $T^{\rm max}_{\rm e7} \leq 7 \cdot {\cal M}_{\rm
-3}^{1/2} \cdot {\cal R}_{\rm -2}^{-1}$. Note that the maximal
electron temperature can be achieved only for FMKs with $v_{\rm 8}
\geq 2.3 \cdot {\cal M}_{\rm -3}^{1/4} \cdot {\cal R}_{\rm
-2}^{-1/2}$.
%and the postshock layer depth ${\cal N}_p \gsim 1.7\cdot
%10^{20}~T_8^{2}~\cms$.

Thermal bremsstrahlung luminosity of the thin postshock plasma
L$_{\rm Tx}$ is $\sim 6\times 10^{31} \cdot {\cal R}_{\rm -2}^3
\cdot n_{\rm a2}^2 \cdot T_{\rm e8}^{1/2}$($\ergs$) (e.g. Rybicki
\& Lightman 1979). The effective average density in the emitting
region is about 3 times that of the ambient (c.f. Wang \&
Chevalier (2002)). The luminosity is consistent with that observed
from the GC sources in the \chan\ catalogue of M03. However, M03
found that spectral modeling of the sources requires extremely hot
temperatures $>$ 25 ${\rm keV}$. That high temperatures can be
expected for relatively small FMKs of ${\cal R}_{\rm -2} \leq$
0.3, propagating through a dense ambient gas of $n_{\rm a} \gsim
10^3 \cmc$. In this case the deceleration time of the FMKs would
be about a few hundred years.

\subsection{The nonthermal emission of FMKs}

Energetic nonthermal particles accelerated by Fermi mechanism in
the MHD collisionless shocks diffuse through the hot postshock
layer and the cold metallic knot, suffering from Coulomb losses
and producing hard X-ray emission both in lines and continuum. The
model of nonthermal emission of FMKs is described in details in
Bykov (2002). The accelerated electron distribution was simulated
using the kinetic description of charged particles interacting
with a strong MHD shock. The X-ray emission is most prominent for
the FMKs in dense ambient media. Energetic particle acceleration
effect is expected to be efficient for strong collisionless shocks
in a magnetized plasma. A significant fraction of the ram pressure
could be transferred into the high energy particles. If that is
the case, the postshock gas pressure drops down, affecting the
reverse shock dynamics and the knot crushing conditions.

The X-ray line emission is due to $K$-shell ionization by
nonthermal particles accelerated by the bow shock and then
propagating through a metal-rich clump. Compact dense knots could
be opaque for some X-ray lines. The optical depth effect due to
resonant line scattering is important and is accounted for in our
model.

%Fig1*******************************

%%%%%%%%%%%%%%%%%%%%%%%%%%%%%%%%%%%%%%%%%%%%%%%%%%%%%%%%%% Fig 1
\begin{figure}
\centering \epsfxsize=85mm
\epsffile{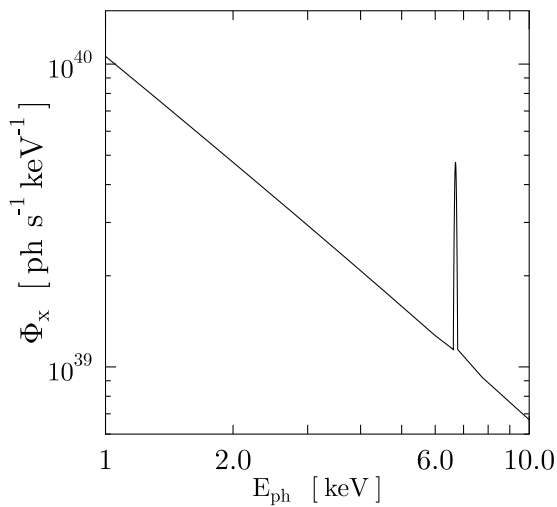} \caption{ Simulated nonthermal emission
spectrum of an FMK of mass ${\cal M}_{\rm -3} \approx 1$, radius
${\cal R}_{\rm -2} = 1$ and velocity $v \approx 5,000 \kms$ in an
ambient medium of n = 100 $\cmc$. } \label{lines}
\end{figure}
%%%%%%%%%%%%%%%%%%%%%%%%%%%%%%%%%%%%%%%%%%%%%%%%%%%%%%%%%%%%%%

In Fig. \ref{lines} we present a simulated spectrum of the
nonthermal X-ray emission from an FMK of 0.01 ${\rm pc}$ size and
mass ${\cal M} \approx 10^{-3} ~\Msun$ dominated by oxygen and
containing $\sim$ 10$^{-4}~\Msun$ of Fe. The knot of velocity $v =
5,000 \kms$ is moving through an ambient gas of $n_{\rm a} = 10^2
\cmc$.

The nonthermal bremsstrahlung spectrum in Fig. \ref{lines} is a
power law of photon index $\Gamma \approx 1.2$ with a prominent Fe
line of the equivalent width $\approx$ 568 ${\rm eV}$. The spectra
with $\Gamma \sim$ 1.2--1.4 are typical in our simulations. The
photon index is in the range $0 \leq \Gamma \leq 1.5$ and depends
on the particle diffusion model inside the FMK and the rate of
nonthermal ions injection (see Bykov 2002).

We simulated the dependence of the nonthermal X-ray luminosity on
the fragment velocity for the same model. The power-law scaling
$L_X \propto v^{\zeta}$ with $\zeta \approx$ 0.6 fits the
simulated data for the FMK velocity range $1.8 < v_8 < 7$.

We also calculated hardness ratios $hr=(h-s)/(h+s)$ for simulated
spectra. The numbers of counts in the high energy band $h$
(4.7--8.0 ${\rm keV}$) and low band $s$ (3.3--4.7 ${\rm keV}$)
were chosen for hard color, and $h$ (3.3--4.7 ${\rm keV}$) and $s$
(2.0--3.3 ${\rm keV}$) for medium color as in M03. For unabsorbed
continuum spectra with the Fe lines simulated with an account of
internal absorbtion we obtained hard color ratios $hr \sim$
0.3--0.4. Without the Fe lines the hard color ratio $hr \lsim$
0.2. The medium color ratio was $\sim$ 0.4.

\section{The ${\rm logN}$--${\rm logS}$ distributions}
The X-ray luminosity of an FMK depends on the fragment velocity,
mass, radius, and the ambient matter. To simulate the ${\rm
log}$N--${\rm log}$S distribution in the homogeneous ambient
medium we should know the FMK velocity and size distributions. An
exact modeling of the distributions would require simulations of
deceleration, expansion and ablation of an FMK as a supersonic,
radiative plasmoid in a strong magnetic field (c.f. Cid-Fernandes
\etal 1996). We consider here a simplified analytic model for
${\rm log}$N--${\rm log}$S distributions.

\subsection{The velocity distribution of FMKs}
Consider SN events (of a rate $\nu_{\rm SN}(\vec{r},t)$) ejecting
$ {\cal{N}}_{\star}$ fragments per event. The  SN rate is a highly
intermittent function with local variations of orders of magnitude
in rich star-forming regions. The differential distribution
function of the FMKs velocities and radii ${\cal
N}(\vec{r},\vec{v},{\cal R},t)$ satisfies

\begin{equation}
\frac{\partial{{\cal N}}}{\partial{t}} +
\dot{\vec{r}}\,\frac{\partial{{\cal N}}}{\partial{\vec{r}}} +
\dot{\vec{v}}\,\frac{\partial{{\cal N}}}{\partial{\vec{v}}} +
\dot{{\cal R}}\,\frac{\partial{{\cal N}}}{\partial{{\cal R}}}=
q(\vec{r},\vec{v},{\cal R},t),\label{eq:VD}
\end{equation}
where the source function $q(\vec{r}, \vec{v},{\cal R},t) =
\nu_{\rm SN}\, {\cal{N}}_{\star}\, \psi(v,{\cal R})$. Here
$\psi(v,{\cal R})$ is the probability distribution of the initial
sizes and velocities for the FMKs. For a ballistically moving FMK
the drag deceleration in the ambient medium is $\dot{\vec{v}}
\propto - \rho_a v\, \vec{v}\, {\cal R}^2$ (for a stretched knot
${\cal R}$ is the transverse radius). The 2D simulations by Klein
\etal (1994) and Wang \& Chevalier (2002) show that the knot
expansion dominate at the early evolution stages, but  stops at $t
> 3\tau_{\rm c}$. We can neglect the drag force at the early knot expansion stage,
while the transverse expansion is not important ($\dot{\cal R}$ =
0) at the later stages where the most efficient drag deceleration
of the expanded FMKs occurs.

We apply the Eq.(\ref{eq:VD}) to the GC region assuming a constant
rate $\nu_{\rm SN}$.
The averaged distribution for $t > 3\tau_{\rm c}$ is
\begin{equation}
{\cal N}(v, {\cal R}) = \int_{v}^{\infty} \frac{\bar{q}(v, {\cal
R})}{\dot{v}}\,\, dv,\label{eq:N}
\end{equation}
where $\bar{q}(v, {\cal R}) \propto a^{-1} q(v,{\cal R}/a)$ and
$a(\chi)$ is an expansion factor. The transverse expansion factor
$a \leq 3$  has a weak dependence on $\chi$  for the range of
parameters relevant to the FMKs (e.g. Klein \etal 1994).

\subsection{Flux distributions}
It is instructive to estimate a power-law index of the flux
distribution function $N(S)$. The index, defined as $\alpha$ =
${\rm log}$N/${\rm log}$S, is constant for a relatively narrow
flux band. We discuss both thermal and nonthermal emission models
described above.

(i) For FMKs of velocity above $v_{\rm 0}({\cal R})$ the postshock
electron temperature could be higher then 8 ${\rm keV}$. Thus,
 the 2-8 ${\rm keV}$ photon flux $S$ studied in the \chan\ sample
depends only weakly on the knot velocity, while $S \propto {\cal
R}^3$. The distribution $\psi$ is not constrained by current
observations and modeling. We simply assume $\psi(v,{\cal R})
\propto {\cal R}^{- \beta} v^{- \eta}$ in a relatively narrow
range of ${\cal R}$. Then, integrating the distribution ${\cal
N}(v, {\cal R})$ over velocity (above the threshold $v_{\rm 0}
\propto {\cal R}^{-1/2}$ to reach the temperature above 8 ${\rm
keV}$) and using $S \propto {\cal R}^3$, we obtain the local index
estimation for the thermal emission $\alpha = -4/3 + \eta/6 -
\beta/3$. The index is consistent with $\alpha = - 1.7 \pm 0.2$
obtained by M03 for the \chan\ GC sample, if the initial
distributions $\psi(v,{\cal R})$ satisfy $-$3.4 $< \eta - 2\beta
<-$ 1.0.

(ii) The nonthermal flux distributions depend on the diffusion
regime of the accelerated particles inside FMKs. The dependence of
the diffusion coefficients on the FMK size is poorly known. Thus
as an illustration we estimate the nonthermal ${\rm log}$N--${\rm
log}$S index $\alpha$ only for an ensemble of FMKs of the same
size ${\cal R}$. We use the relation $S \propto v^{\zeta}$
discussed above to get $\alpha = - \eta/\zeta - 1$. Since the
simulated index $\zeta \approx 0.6$, the ${\rm log}$N--${\rm
log}$S distribution index $\alpha$ would be consistent with the
\chan\ sample data for the initial velocity distributions with
$\eta < 0.8$. The value of $\eta \lsim 1$ agrees with the velocity
distribution of FMKs simulated by Kifonidis \etal (2003).

\section{Conclusions}
Any SN event in the GC region is expected to produce many hundreds
of fast moving fragments of  mass $\sim$ 10$^ {-3} \Msun$. If a SN
ejects ${\cal{N}}_{\star} \sim$ 500 FMKs of velocity 5,000 $\kms$,
 the FMKs would carry about 10\% of the SN kinetic energy.
Three SNRs during the last 1,000 ${\rm years}$ could produce an
ensemble of more than 1,500 FMKs in the GC region providing a
normalization of the ${\rm log}$N--${\rm log}$S consistent with
the \chan\ data. The filamentary features apparent in the \chan\
images of the GC region could be also relevant to the SN activity.

Fast moving SN ejecta fragments interacting with the dense GC
environment provide a fast conversion of SN kinetic energy into
the IR and the X-ray emission.  The X-ray spectra of the
fragments (with L$_{\rm x} \leq 10^{33} \ergs$ per FMK) are hard.
They contain thermal and power law components with possible Fe
lines. Both 6.7 and 6.4 ${\rm keV}$ lines are expected, depending
on the FMK ionization structure and the relative strength of the
nonthermal component. The total equivalent width is about 500 -
600 ${\rm eV}$ for a Fe mass of $\sim$ 10$^{-4}~\Msun$. The hard
color ratio $hr$ is $\lsim$ 0.2 without Fe lines, while it is
$\sim$ 0.3--0.4 for a Fe mass of $\sim$ 10$^{-4}~\Msun$. The
medium color $hr$ is $\sim$ 0.4.

The spatial density of the FMKs brighter than some fixed
luminosity  $\cal{L}_{\rm 0}$ scales $\propto n_{\rm
a}^{\gamma}(r)$ with the ambient gas density. This is because
$L_{\rm x}\propto n_{\rm a}^{\nu}$ for a single FMK, while ${\cal
N} \propto n_{\rm a}^{-1}$ from Eq.(\ref{eq:N}). The index $\gamma
= -\nu (\alpha +1) - 1$ depends on the ${\rm log}$N--${\rm log}$S
distribution. For the thermal emission model $\nu = 2$. That
implies a global decrease in the spatial density of the sources
$\propto n_{\rm a}^{0.4}(r)$ (for $\alpha$ = $-$ 1.7) away from
Srg A$^{\ast}$, though strong local fluctuations could appear. The
apparent surface density of the FMKs would depend on the symmetry
of the SN distribution (e.g. disk or spherical). The global
distribution of the FMKs is expected to be an ensemble of coherent
shell- and jet-like structures of a few $arcmin$-scale.

\begin{acknowledgements}
I thank the referees for helpful comments. The work was supported
by RBRF 03-02-17433, 01-02-16654.

\end{acknowledgements}

%\end{flushleft}
\end{document}